\documentclass[reprint,superscriptaddress,preprintnumbers,amsmath,amssymb,aps,prd,tightenlines,longbibliography,balancelastpage]{revtex4-2}

\usepackage{graphicx}
\usepackage[dvipsnames]{xcolor}
\usepackage[sort&compress]{natbib}
\usepackage{amsmath,amssymb,bm,bbm,slashed,subdepth}
\usepackage{xr-hyper}
\usepackage[colorlinks=true
,urlcolor=blue
,anchorcolor=blue
,citecolor=blue
,filecolor=blue
,linkcolor=red
,menucolor=blue
,linktocpage=true
,pdfproducer=medialab
,pdfa=true
]{hyperref}
\usepackage{cleveref}
\usepackage{enumerate}
\usepackage{epsfig, subfigure}
\usepackage{setspace}
\usepackage{booktabs, tabularx}
\usepackage{units}
\usepackage{placeins}
\usepackage{multirow}
\usepackage{mathtools}
\usepackage[normalem]{ulem}
\usepackage{orcidlink} 
\usepackage[left,modulo]{lineno}
\modulolinenumbers[1]

\makeatletter
\newcounter{savesection}
\newcounter{apdxsection}
\renewcommand\appendix{\par
  \setcounter{savesection}{\value{section}}%
  \setcounter{section}{\value{apdxsection}}%
  \setcounter{subsection}{0}%
  \gdef\thesection{\@Alph\c@section}}
\newcommand\unappendix{\par
  \setcounter{apdxsection}{\value{section}}%
  \setcounter{section}{\value{savesection}}%
  \setcounter{subsection}{0}%
  \gdef\thesection{\@arabic\c@section}}
\makeatother

\begin{document}

\title{Discovering the Higgsino at CTAO-North within the Decade}

\author{Shotaro Abe\,\orcidlink{0000-0001-7250-3596}}
\affiliation{Institute for Cosmic Ray Research, University of Tokyo, Kashiwa, Chiba 277-8582, Japan}

\author{Tomohiro Inada\,\orcidlink{0000-0002-6923-9314}}
\affiliation{Kyushu University, Nishi-ku, 819-0395 Fukuoka, Japan}
\affiliation{CERN, CH-1211 Geneva 23, Switzerland}
\affiliation{Institute for Cosmic Ray Research, University of Tokyo, Kashiwa, Chiba 277-8582, Japan}

\author{Emmanuel Moulin\,\orcidlink{0000-0003-4007-0145}}
\affiliation{Irfu, CEA Saclay, Université Paris-Saclay, F-91191 Gif-sur-Yvette, France}

\author{\mbox{Nicholas L. Rodd\,\orcidlink{0000-0003-3472-7606}}}
\affiliation{Berkeley Center for Theoretical Physics, University of California, Berkeley, CA 94720, U.S.A.}
\affiliation{Theoretical Physics Group, Lawrence Berkeley National Laboratory, Berkeley, CA 94720, U.S.A.}

\author{Benjamin R. Safdi\,\orcidlink{0000-0001-9531-1319}}
\affiliation{Berkeley Center for Theoretical Physics, University of California, Berkeley, CA 94720, U.S.A.}
\affiliation{Theoretical Physics Group, Lawrence Berkeley National Laboratory, Berkeley, CA 94720, U.S.A.}

\author{Weishuang Linda Xu\,\orcidlink{0000-0001-5170-409X}}
\affiliation{Berkeley Center for Theoretical Physics, University of California, Berkeley, CA 94720, U.S.A.}
\affiliation{Theoretical Physics Group, Lawrence Berkeley National Laboratory, Berkeley, CA 94720, U.S.A.}
\affiliation{Kavli Institute for Particle Astrophysics \& Cosmology, Stanford University, Stanford, CA 94305, U.S.A.}
\affiliation{Particle Theory Group, SLAC National Accelerator Laboratory, Stanford, CA 94305, U.S.A.}

\begin{abstract}
We demonstrate that higgsino dark matter (DM) could be discovered within the next few years using the  Cherenkov Telescope Array Observatory's soon-to-be-operational northern site (CTAO-North).
A 1.1\,TeV thermal higgsino is a highly motivated yet untested model of DM.
Despite its strong theoretical motivation in supersymmetry and beyond, the higgsino is notoriously difficult to detect; it lies deep within the neutrino fog of direct detection experiments and could pose a challenge even for a future muon collider.
We show that, in contrast, higgsino detection could be possible within this decade with CTAO-North in La Palma, Spain.
The Galactic Center is the region where the dominant DM annihilation signature emerges, but it only barely rises above the horizon at the CTAO-North site.
However, we project that this challenge can be overcome with large-zenith-angle observations at the northern site, enabling the conclusive detection of a higgsino signal by 2030 for a range of DM density profiles in the inner Galaxy.
\end{abstract} 

\maketitle

The higgsino may be the last chances for nature to realize a minimal weakly interacting massive particle (WIMP) as the dark matter (DM) of our Universe.
A nearly pure higgsino commonly emerges as the lightest supersymmetric particle in supersymmetric extensions of the Standard Model~\cite{Wells:2003tf,Giudice:2004tc,Arkani-Hamed:2004ymt,Hall:2011jd,Arvanitaki:2012ps,Arkani-Hamed:2012fhg,Co:2021ion}.
Realizing the WIMP miracle, the higgsino can be produced as a thermal relic of the early Universe with the correct DM abundance when $m_\chi \simeq 1.1$\,TeV~\cite{Bottaro:2022one}; we define the higgsino with this mass as the thermal higgsino.
With other WIMP candidates such as the wino facing increasingly strong tension from null indirect detection searches~\cite{Fan:2013faa,Cohen:2013ama,Rinchiuso:2018ajn,Rodd:2024qsi}, the higgsino is arguably the last path to a strongly motivated realization of the minimal DM paradigm~\cite{Cirelli:2005uq}.

\begin{figure}[!t]
\centering
\includegraphics[width=\linewidth]{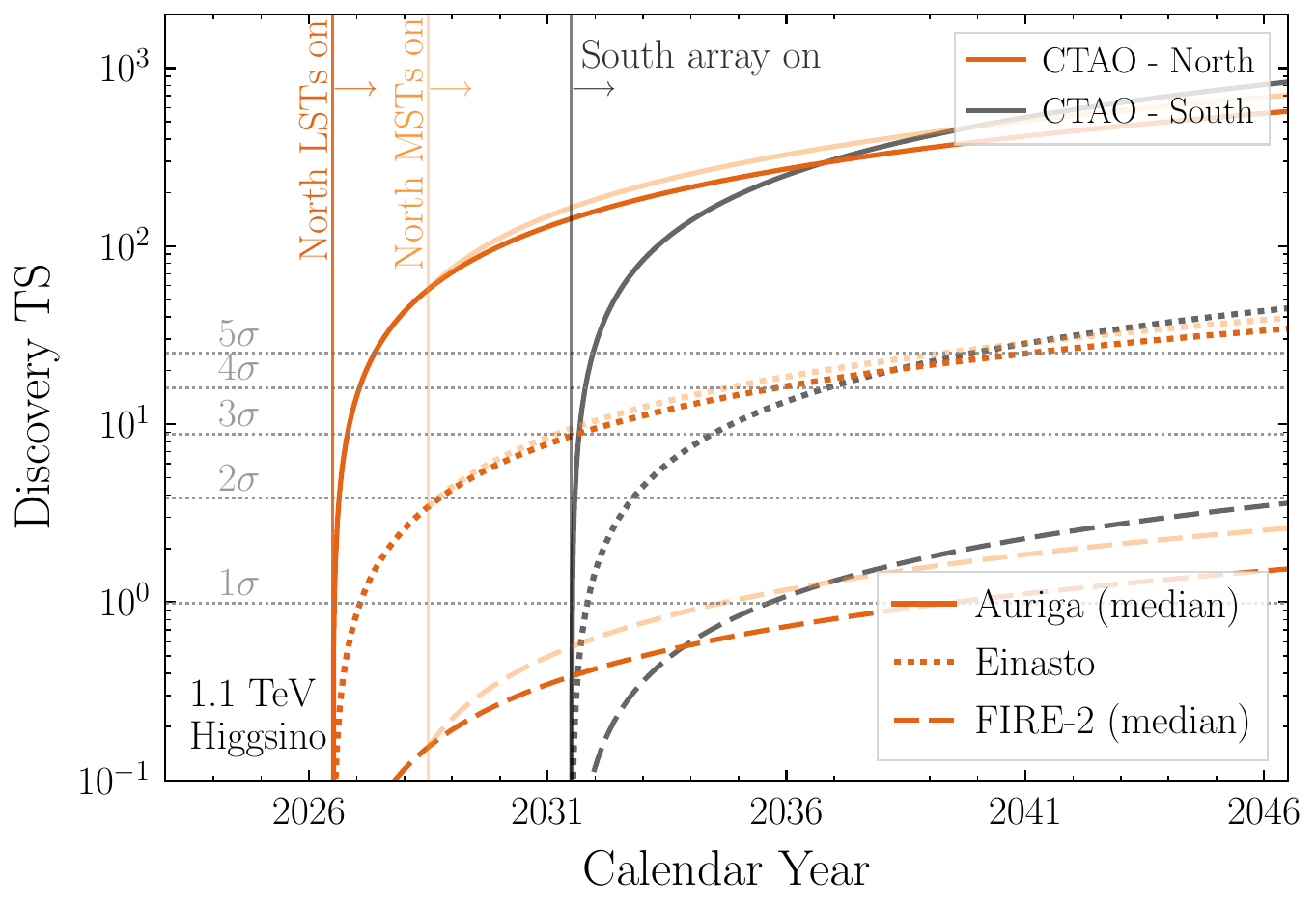}
\vspace{-0.8cm}
\caption{The expected discovery test statistic (TS) for the 1.1\,TeV higgsino DM model, including both the line and continuum contributions, at CTAO-North (assumed to start mid 2026--labeled ``North LSTs on'') and CTAO-South in the Alpha Configuration (assumed start mid 2031) as a function of time.
No intermediate-array configurations for CTAO-South are considered.
For CTAO-North we show sensitivity from the four LSTs alone (orange) as well as four LSTs + nine MSTs starting in 2028 (lightly shaded orange).
We assume a data collection rate of 100\,hrs/year in the Galactic Center region around 60$^{\circ}$ in zenith angle.
We illustrate the discovery prospects across a wide span of underlying DM profile possibilities, displaying the median case from the Auriga and FIRE-2 hydro simulations, along with the DM-only-based Einasto profile.
}
\vspace{-0.5cm}
\label{fig:TS_over_time}
\end{figure}

Higgsino DM is a key science driver of many next generation experiments in particle physics, ranging from future DM direct detection efforts~\cite{Akerib:2022ort,Bloch:2024suj} to next generation colliders~\cite{Low:2014cba,Capdevilla:2021fmj,Accettura:2023ked,Saito:2019rtg,EuropeanStrategyforParticlePhysicsPreparatoryGroup:2019qin,Canepa:2020ntc,Black:2022cth,Fukuda:2023yui,Franceschini:2022sxc,Narain:2022qud,Aime:2022flm,MuonCollider:2022xlm,Mahbubani:2017gjh,Capdevilla:2024bwt,Delgado:2020url} to precision experiments such as those searching for an electron electric dipole moment~\cite{Giudice:2005rz,Krall:2017xij,Cesarotti:2018huy}.
Nonetheless, thermal higgsino DM is notoriously difficult to probe with terrestrial experiments.
In its most minimal form the direct detection cross-section is well below the neutrino floor and thus out of reach of any planned near- or mid-term effort~\cite{Hisano:2011cs,Hill:2013hoa,Hill:2014yka,Hill:2014yxa,Chen:2019gtm}, though if the higgsino strongly mixes with other neutralino states it could be more detectable~\cite{Martin:2024pxx,Graham:2024syw}.
As an example of the difficulty of higgsino detection, the thermal higgsino is only just within reach of proposed future 10\,TeV muon colliders and the Future-Circular Collider (FCC)-hh; accounting roughly for the full expected luminosity of these colliders the thermal higgsino may be detected with around $\sim$5$\sigma$ significance~\cite{Han:2020uak,Capdevilla:2021fmj}. 
The thermal higgsino is well out of reach of the Large Hadron Collider (LHC), including in all proposed future runs of the LHC such as high luminosity and high energy runs.

In contrast to terrestrial searches, indirect searches for thermal higgsino annihilation appear more promising in the coming decades.
Higgsinos may annihilate in the center of the Galaxy to $W^+W^-$ and $ZZ$ at tree level and to $\gamma Z$ and $\gamma \gamma$ at one loop~\cite{Beneke:2022eci}.
The tree-level annihilation products shower to create a broad spectrum of gamma-rays below a TeV, while the one-loop processes create a narrow spectral endpoint, which in practice is broadened primarily by the finite energy resolution of the gamma-ray detectors.
Searches using data from the {\it Fermi} Large Area Telescope (LAT) to-date have leading sensitivity to higgsino DM annihilation, potentially even seeing the first signs of an emerging signal~\cite{Dessert:2022evk}.
Of course, searches for continuum emission are complicated by the other confounding astrophysical backgrounds.
In contrast, there is no known, standard astrophysical process that is capable of making a narrow spectral feature at energies $\sim$1\,TeV.
Thus finding a $\sim$1.1\,TeV gamma-ray line towards the Galactic Center would be a smoking gun pointing to higgsino annihilation.

The Cherenkov Telescope Array Observatory (CTAO) is the next generation of ground-based observatories designed to explore high-energy gamma-ray astronomy~\cite{CTAConsortium:2010umy,CTAConsortium:2017dvg}.
Spanning an energy range from 20\,GeV to 300\,TeV, CTAO employs three types of telescopes -- Large-Sized Telescopes (LSTs), Medium-Sized Telescopes (MSTs), and Small-Sized Telescopes (SSTs) -- in order to achieve such energy coverage.
The telescopes are distributed across two sites in order to cover the full sky: CTAO-South, located in Paranal, Chile, and CTAO-North, situated at the Roque de los Muchachos Observatory on La Palma, Canary Islands.

Statements regarding the discovery prospects for particle DM leveraging the relic DM abundance often hinge on an assumption of the Milky Way's DM distribution.
Modeling the DM density profile in the inner Galaxy is notoriously difficult because the gravitational potential in this region is dominated by baryons, {\it e.g.} Ref.~\cite{Hussein:2025xwm}.
We return to this point later in this work.
For the moment, we note that current space- and ground-based gamma-ray telescopes, such as {\it Fermi}-LAT, MAGIC and H.E.S.S., do not have the necessary sensitivity to resolve the line-like signal from higgsino annihilation except for the most optimistic DM distributions (see, {\it e.g.},~\cite{MAGIC:2022acl,HESS:2018cbt,Fermi-LAT:2015kyq}).
On the other hand, it has been established that the upcoming CTAO-South site in Chile will achieve the necessary sensitivity to conclusively discover higgsino DM for all but the most pessimistic Galactic DM profiles~\cite{Rinchiuso:2020skh,Rodd:2024qsi}.

The first telescopes for CTAO-South are expected to be delivered in 2026 and begin collecting scientific data within a couple of years~\cite{ERIC}.
Here we assume that the full Alpha Configuration will not begin until 2031, although a partial southern array may be available sooner.
For CTAO-North, LST-1 in La Palma has already achieved first light~\cite{CTA-LSTProject:2023haa}.
An initial detector configuration, with four LSTs, is expected to become operational in 2026.

Although the northern site is poised to be operating sooner, CTAO-South is the conventional focus of DM annihilation searches as the Galactic Center rises to much lower zenith angles in the southern hemisphere~\cite{CTAO:2024wvb}.
In particular, the Galactic Center rises to $<$10$^\circ$ zenith angle at CTAO-South, compared to $\sim$58$^\circ$ at CTAO-North.
Nevertheless, is has been demonstrated that Imaging Atmospheric Cherenkov Telescopes (IACTs), such as CTAO, can reliably collect data near the horizon.
While this comes with drawbacks like increased background rates and a higher energy threshold due to dimmer shower images, it also offers the benefit of a larger effective area compared to observations at low zenith angles.
For example, the MAGIC telescope~\cite{Albert:2005kh}, which is also located in La Palma, has observed the Galactic Center for over 200\,hrs to perform a search for DM annihilation including line-like final states~\cite{MAGIC:2020kxa, MAGIC:2022acl}.
For the first time, in this Letter we consider the ability of CTAO-North to use large-zenith-angle observations of the Galactic Center to search for signals of higgsino annihilation. 
Our conclusions are displayed in Fig.~\ref{fig:TS_over_time}: the higgsino may be discovered or disfavored within a few years using upcoming data from CTAO-North, depending on the DM profile near the Galactic Center.

\vspace{0.2cm}
\noindent {\bf Higgsino Annihilation.}
%
The higgsino is nearly fully specified by its electroweak representation and mass. Thus, one can precisely compute both its annihilation rate and the photon spectrum that emerges per annihilation.
We use \texttt{DM$\gamma$Spec}~\cite{Beneke:2022eci} to determine the spectrum; here, we briefly explain the physics involved.
There are three relevant contributions to the spectrum that one needs to consider for the higgsino: the line, endpoint, and continuum.
Let us discuss the physical origin and importance of each.

Line photons emerge from exclusively two body final states such as $\gamma \gamma$ and $\gamma Z$.
(For the thermal higgsino mass the photon energies from these two distinct final states are within 0.2\% of one another.)
Although the DM is electrically neutral, these two body final states are reached at loop level and are subject to Sommerfeld enhancement~\cite{Hisano:2003ec,Hisano:2004ds,Cirelli:2007xd,Arkani-Hamed:2008hhe}.
This enhancement arises as the higgsino experiences an effectively long-range force due to electroweak interactions, which enhances their annihilation cross-section at low velocities.

The endpoint contribution arises from three or more body final sates, which we collectively denote by $\gamma+X$.
What is important about these states is that we demand that the photon energy $E$ is very close to the DM mass, in particular $1-E/m_{\chi} \ll 1$.
Given that the CTAO energy resolution is at minimum $\sim$5\% (see Fig.~\ref{fig:comp}), once we apply the finite energy resolution of the detector, these photons are indistinguishable from those of the $\gamma\gamma$ and $\gamma Z$ final states.
An example of such a three-body final state is $\gamma W^+ W^-$, where if the two $W$-bosons are sufficiently collinear, then the photon can have $E \sim m_{\chi}$.
The contribution from these endpoint photons can be enhanced by large logarithms that are resummed using effective field theory techniques, as has been broadly explored for electroweak DM candidates (see Refs.~\cite{Baumgart:2014vma,Bauer:2014ula,Ovanesyan:2014fwa,Baumgart:2014saa,Baumgart:2015bpa,Ovanesyan:2016vkk,Baumgart:2017nsr,Baumgart:2018yed,Beneke:2018ssm,Beneke:2019vhz,Beneke:2022eci,Beneke:2022pij,Baumgart:2023pwn}).
Ultimately, these effects make an ${\cal O}(1)$ correction to the expected intensity of the line-like signal at $E  = m_\chi$.

Lastly, there are the lower-energy gamma-ray contributions from the tree-level $ZZ$ and $WW$ final states.
The $W$ and $Z$ bosons are unstable and when they shower and decay they produce a continuum spectrum of photons with energies $E \ll m_{\chi}$.
A large fraction of these photons can fall in the CTAO energy band depending on the energy range of the analysis.
These effects are incorporated in \texttt{DM$\gamma$Spec} using the computation from \texttt{PPPC4DMID}~\cite{Cirelli:2010xx}.
(See Ref.~\cite{Jager:2023njk} for additional considerations for sub-TeV higgsinos.)

\vspace{0.2cm}
\noindent {\bf CTAO-North Instrument Response.}
%
The observational performance of IACTs strongly depends on the zenith angle.
At CTAO-North's location on La Palma, the Galactic Center culminates at a zenith angle of 58$^\circ$, which is challenging for traditional IACT observations.
Larger zenith angles result in a greater distance to the shower and a geometrically wider light pool, leading to an enhanced effective area---even by an order of magnitude.
Yet the Cherenkov light is further attenuated relative to low-zenith angle observations, leading to an increase of the telescope's energy threshold, as we discuss further below (see, {\it e.g.}, Ref.~\cite{MAGIC:2014zas}).

\begin{figure}
\centering
\includegraphics[width=\linewidth]{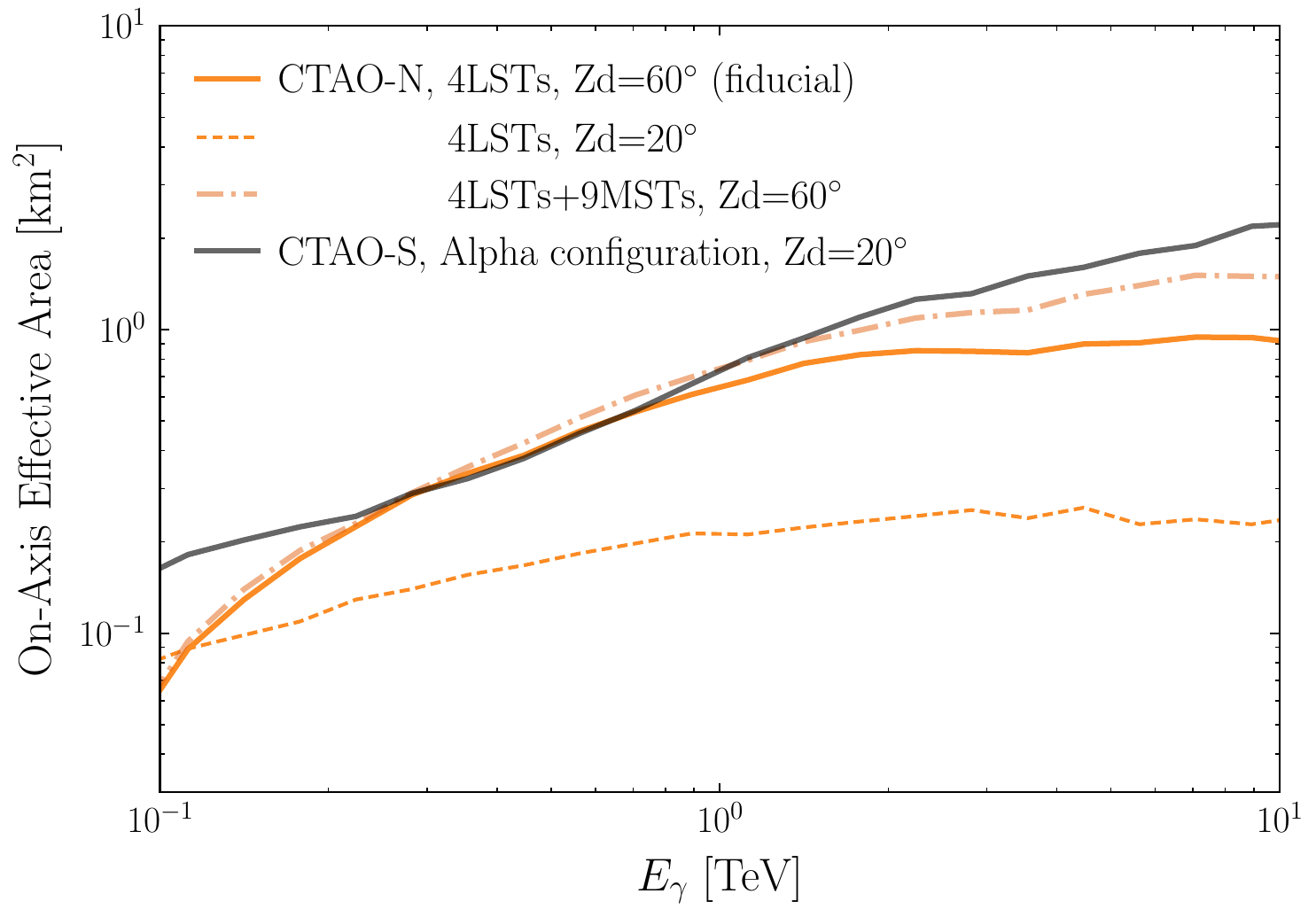}
\vspace{-0.8cm}
\caption{The on-axis effective area as a function of the photon energy $E_\gamma$ for different combinations of CTAO-North and South configurations \cite{cherenkov_telescope_array_observatory_2021_5499840}.
At 1\,TeV the effective area of the LST-only northern array at large zenith angle (Zd) is comparable to CTAO-South.
}
\vspace{-0.5cm}
\label{fig:Aeff}
\end{figure}

For this study, we assume an average zenith angle of 60$^\circ$ and further that the data is collected via stereo observations involving all available LSTs and also MSTs once they become operational.
We use Instrument Response Functions (IRFs) tailored for large zenith angle observations from Monte Carlo simulations using the publicly available \texttt{prod5-v0.1} library~\cite{CTAO_2021_5499840}.  
The IRFs comprise the energy/offset-dependent effective area, angular resolution, and energy resolution for the typical zenith angle.
In Fig.~\ref{fig:Aeff} we compare the effective area with different zenith angles (20$^{\circ}$ and 60$^{\circ}$) and telescope array setups at both CTAO sites.
We assume telescope configurations for the North and South based on the latest, official CTAO plan~\cite{CTA_performance}.
For CTAO-South we adopt the Alpha Configuration, which consists of 14 MSTs and 37 SSTs.
For CTAO-North, we instead consider a two-stage telescope configuration, where first there are 4 active LSTs that are then supplemented by 9 MSTs.
For CTAO-North we show the effective area at small- and large-zenith angles ($20^\circ$ and 60$^\circ$, respectively), while for CTAO-South we only show small zenith angles because the Galactic Center rises well above the horizon.
In our projections in Fig.~\ref{fig:TS_over_time} we assume that the 4 LSTs for CTAO-North begin collecting science data in 2026, with the additional 9 MSTs turning on optimistically two years later in 2028.
We assume for this work that the CTAO-South Alpha Configuration begins collecting science data in 2031.

A notable feature of Fig.~\ref{fig:Aeff} is that at TeV energies moving from small to large zenith angles increases the effective area by almost an order of magnitude.
This increase is the primary driver of the higgsino reach projected in this work, as the increase of the energy threshold to a few hundreds of GeV is not impactful on the line signals at TeV energies.
Note that the 9 MSTs do not appreciably increase the on-axis effective area at 1\,TeV, but they do increase the effective area off-axis, as we show in Fig.~\ref{fig:comp}.
The MSTs are therefore more impactful in constraining scenarios where the DM distribution is more cored, for which the signal falls off more slowly with distance from the Galactic Center.
In fact, at energies above hundreds of GeV as shown in Fig.~\ref{fig:Aeff}, the large-zenith-angle CTAO-North effective area is comparable to that of small-zenith-angle observations with CTAO-South, despite CTAO-South having more telescopes than the northern site.
However, as discussed next, the trade off is that large-zenith-angle observations are accompanied by decreased energy resolution and background rejection efficiency (see Fig.~\ref{fig:comp}).

Observations at large zenith angles typically involve greater systematic uncertainties when compared to those at small zenith angles.
For instance, previous studies using the MAGIC telescopes have reported systematic uncertainties, for large (small) zenith angles, of approximately 15\% (10\%) for the absolute energy scale and 20\% (15\%) for flux normalization~\cite{Ahnen:2016crz, 2015PhDT176F, MAGIC:2020kxa, MAGIC:2022acl}.
Systematics related to the absolute energy scale and overall flux normalization are straightforwardly mitigated; {\it e.g.}, a nuisance parameter can be assigned to the overall flux normalization as in Ref.~\cite{Montanari:2022buj,Rodd:2024qsi}.
On the other hand, it is also important to account for the energy-dependent observation-to-observation scatter in the cosmic-ray rejection efficiency.
To quantify the impact of such a systematic, one in general needs to construct a spectral model for the residual cosmic-ray background from an ensemble of OFF observations, made at the zenith angle of interest but pointings away from the Galactic Center, and study how well this model describes individual realizations of that ensemble.
We postpone detailed studies of systematic uncertainties  until CTAO data becomes available; in this work, we concentrate on idealized analysis frameworks to showcase the maximal science capabilities of the upcoming instruments.

\begin{figure}[!t]
\centering
\includegraphics[width=\linewidth]{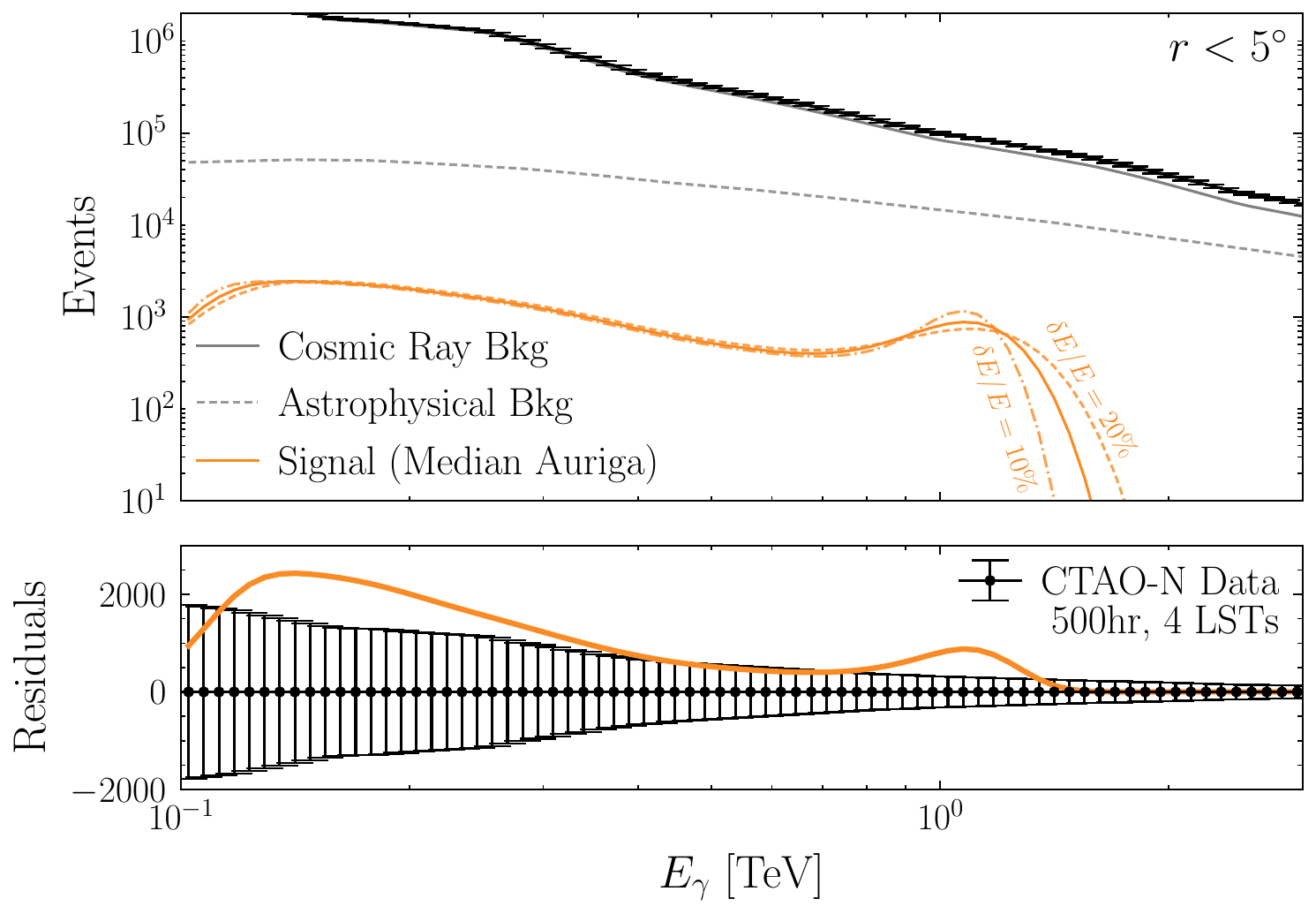}
\vspace{-0.8cm}
\caption{The expected background and signal counts within $5^\circ$ of the Galactic Center for 500\,hrs total of observation time with CTAO-North LSTs only.
We show the total expected counts and associated statistical uncertainties, per energy bin, for the Asimov data set, along with the expected signal and background contributions from misidentified cosmic rays and astrophysical backgrounds, where for the signal model we use the median Auriga profile (see Fig.~\ref{fig:J_facs}).
Note that the signal model has been convolved with the expected instrument response, which broadens the narrow end-point line.
In addition to the benchmark assumption of 15\% energy resolution, we illustrate how the signal profile is altered for 10\% and 20\% energy resolutions.
The bottom panel demonstrates how the signal would emerge in the residual dataset that follows from the subtraction of the best fit background model.
}
\vspace{-0.5cm}
\label{fig:Signal_Background_spectra}
\end{figure}

\vspace{0.2cm}
\noindent {\bf Discovery Prospects at CTAO-North.}
%
At maximum, CTAO-North could collect roughly 250\,hrs/year of dark time data toward the Galactic Center from 58$^{\circ}$ to 70$^{\circ}$ in zenith angles~\cite{TeVCat}.
We assume for definiteness 100\,hrs/year of data is collected in practice.
We further assume that the exposure time is split equally into four pointing locations slightly offset from the Galactic Center: $[\ell, b] = \{ [\pm  0.42^\circ, \pm 0.42^\circ]\}$, with the four locations corresponding to the four sign combinations.
In reality, the pointing locations at the Galactic Center will be optimized to maximize the science impact across a variety of science targets, given the large number of gamma-ray sources in this region; our simple pointing pattern is based on the assumption of the LST-1 monoscopic operation~\cite{CTALSTProject:2023wzu} and does not attempt to make such an optimization~\cite{CTA:2020qlo}.
The LSTs are designed to optically cover a field of view (FOV) with a diameter of 4.3$^\circ$ (see Fig.~\ref{fig:comp}).

A mock exposure map is provided in Fig.~\ref{fig:exp_map}.
The FOV is larger when the MSTs are included (diameter $\sim$8$^\circ$), and larger still for the small zenith angle CTAO-South configuration.
In light of these changes in FOV, for projections involving the full-array realizations of CTAO North and South, we assume pointing locations offset at $1.2^\circ$ as opposed to $0.6^\circ$ from the Galactic Center (see Fig.~\ref{fig:exp_map}).
For our idealized projections, this does not substantially impact our results, but we note that these choices may be important in realistic observing scenarios.
The energy resolution is taken to be $\sim$15\% (see the End Matter for illustrations of the IRFs).
We assume a cosmic-ray-induced background rate derived from dedicated Monte Carlo simulations  given by the {\tt prod5-v0.1} library~\cite{cherenkov_telescope_array_observatory_2021_5499840}.
An additional background component is the astrophysical flux from the Galaxy, which we account for with the {\it Fermi} {\tt gll\_iem\_v07 ({\texttt p8r3})} diffuse model; we extrapolate above 2\,TeV using a power-law as in Ref.~\cite{Rodd:2024qsi}.
Note that the {\texttt p8r3} model is a good description of the gamma-ray sky at TeV energies in the inner Galaxy, as validated through {\it Fermi} Large Area Telescope (LAT) data (see Ref.~\cite{Rodd:2024qsi} for a discussion).
These two contributions, dominating the counts on a detector, are compared to a putative signal flux in Fig.~\ref{fig:Signal_Background_spectra}.

\begin{figure}[!t]
\centering
\includegraphics[width=\linewidth]{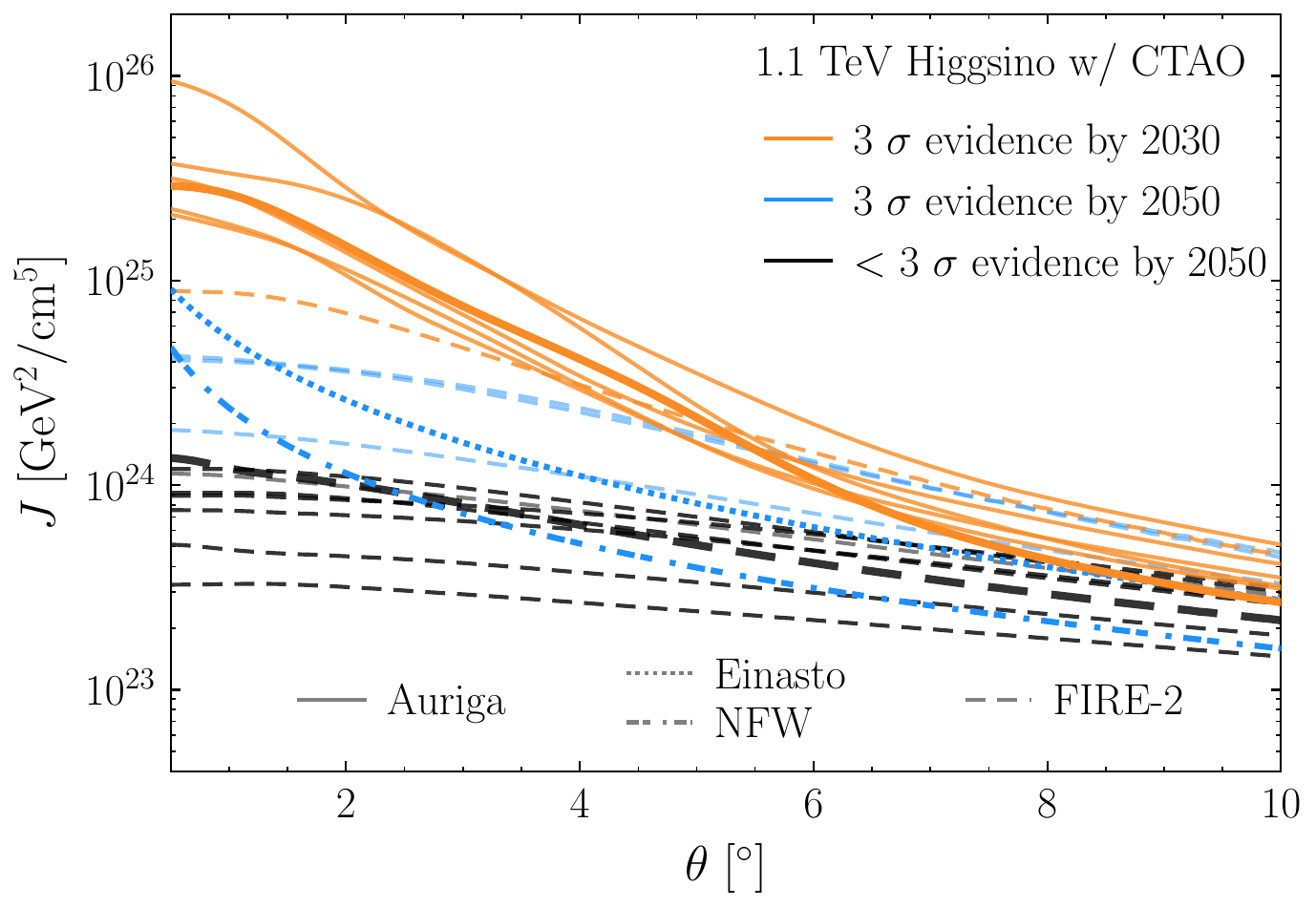}
\vspace{-0.8cm}
\caption{The expected $J$-factors (see End Matter)  as a function of the angle $\theta$ away from the Galactic Center for a number of the spherically-symmetric DM density profiles we consider.
We show the DM-only-motivated Einasto (dotted) and NFW (dot-dashed) models along with profiles from hydrodynamic simulations of Milky Way analogue galaxies, which crucially account for baryonic driven effects in the inner Galaxy, specifically from Auriga (solid) and FIRE-2 (dashed).
The median profiles from each of these suites have been highlighted in bold, and these along with the Einasto profile are used to produce the results in Fig.~\ref{fig:TS_over_time}.
Additionally, we color code the $J$-factor profiles by those which are expected to lead to higgsino evidence of at least 3$\sigma$ with CTAO by 2030 (North only) and 2050 (North and South).}
\vspace{-0.5cm}
\label{fig:J_facs}
\end{figure}

Since baryonic physics dominates the gravitational potential at the center of the Milky Way, it may well be that common ansatzes such as the Einasto profile~\cite{1965TrAlm...5...87E}, empirically motivated from DM-only simulations, do not realistically reflect the underlying DM distribution of our Galaxy. 
To bracket the current wide range of uncertainty on the DM content in the inner Galaxy we use two suites of hydro-intensive high-resolution simulations, Auriga~\cite{Grand:2016mgo,Grand:2024xnm} and FIRE-2~\cite{McKeown:2021sob}, each of which contains a set of Milky Way-like galaxies with candidate profiles.
In our analysis we consider 6 galaxies from Auriga, 12 galaxies from FIRE-2, as well as the DM-only NFW and Einasto profiles, all normalized such that the density at radii of 8.3\,kpc is 0.38\,GeV/cm$^3$.
The $J$-factors of these profiles, shown in Fig.~\ref{fig:J_facs}, are defined as the integral of the DM density squared along the line of sight (see the End Matter), and illustrates the range of signal sizes we expect from DM annihilation.
As is clear from Fig.~\ref{fig:TS_over_time}, the median Auriga prediction for the amount of DM in the inner Galaxy is significantly larger than that of FIRE-2.
These predictions roughly bracket our current uncertainty on the $J$-factor; the differences arise from an interplay of baryons near the Galactic Center generating an adiabatic contraction of the DM halo and strong baryonic feedback -- arising, for instance, from supernovae -- that disperse the baryons from the center, partially dragging DM with them.
For a detailed discussion of these points see Ref.~\cite{Hussein:2025xwm}.
(Note previous indirect detection efforts have occasionally assumed the cored Burkert profile ({\it e.g.}, Ref.~\cite{MAGIC:2022acl}), which leads to a lower density than any of the DM profiles considered here, though modern hydrodynamic simulations are not consistent with such large cores~\cite{Hussein:2025xwm}.)

To project sensitivity to a DM annihilation signal we use the following idealized analysis scheme, which is based off of that discussed in Ref.~\cite{Rodd:2024qsi} for CTAO-South.
We bin the Asimov ({\it i.e.}, mean, expected) data~\cite{Cowan:2010js} in 100 logarithmically-spaced energy bins between 100\,GeV and 10\,TeV.
For concreteness, we bin the data from all observations spatially in five concentric annuli around the Galactic Center of radius $1^\circ$ each, though our results are not strongly sensitive to this choice.
We analyze the data over energy bins in each annulus separately. We model the spectral data using two background spectral templates, one for cosmic-ray-induced background emission and one for astrophysical backgrounds, in addition to the signal spectral template.
Point-source and other extended components are not included in this analysis as they are negligible compared to the diffuse emission in all annular bins~\cite{Rodd:2024qsi}.
Note that we include all energy bins in our analyses, so that both the annihilation end-point and continuum contribute to the signal model.
As discussed in the End Matter, the line plus end-point and continuum components contribute roughly equally to the discovery potential, though the line plus end-point component may be easier to identify in a realistic analysis given its narrow spectral extent.
In Fig.~\ref{fig:Signal_Background_spectra} we illustrate the spectral templates for CTAO-North, with LSTs only, summed over all annuli ($r < 5^\circ$) along with the Asimov data and 1$\sigma$ statistical error bars, assuming 500\,hrs of exposure time.
The cosmic-ray-induced background dominates in all but the inner most radii~\cite{Rodd:2024qsi}.

To show the maximal science reach of the instruments, we assume a perfect model for the background components, with no free parameters; in reality, a realistic analysis will almost certainly have nuisance parameters that will degrade the sensitivity relative to this estimate~\cite{Rodd:2024qsi}.
In order to ensure the null model is nested within the signal model, we multiply the signal normalization by a parameter $\mu$, with $\mu = 0$ giving the null hypothesis and $\mu = 1$ the expectation for the thermal higgsino, at fixed higgsino mass.
We compute the likelihood for the signal model parameter $\mu$ in each annulus and then join the likelihoods together across the annuli to construct our final statistic used to probe the higgsino model.
We then assume Wilks' theorem and use standard frequentist methods to project the expected discovery significance under the signal hypothesis ($\mu = 1$).
See {\it e.g.} Ref.~\cite{Safdi:2022xkm} for more details.

Implementing the procedure above, in Fig.~\ref{fig:TS_over_time} we display the discovery potential of CTAO North and South as a function of calendar year.
We quantify the discovery potential through the discovery test statistic (TS) under the signal hypothesis, as described in the End Matter; the discovery significance in terms of number of ``$\sigma$'' is approximately $\sqrt{\rm TS}$.
For the northern site, we assume that the full LST array will begin data collection in 2026, and that the MST array could contribute two years later.
For the southern site, we assume that the Alpha Configuration begins data taking in 2031 and as for CTAO-North we use an optimisitc data collection rate of 100\,hrs per year.
The CTAO-South instrument response and assumed cosmic-ray-induced background rates are illustrated in the End Matter.

We project the discovery TS as a function of time for a variety of assumptions for the DM density profile.
We show the canonical DM-only Einasto profile and the medians of the ensemble of Auriga and FIRE-2 profiles, with medians computed with respect to the ensemble of discovery TSs across the simulations.
We find that the Auriga median discovery TSs are over two orders of magnitude larger than the those for FIRE-2, with the Einasto profile in between.
In the Auriga $J$-factor scenario, CTAO-North is able to detect higgsino DM with over $5$$\sigma$ significance by 2028.
The median FIRE-2 model is far more challenging and would require decades of data collection with CTAO-South or further improvements to the instruments, although the thermal higgsino could still be reached well before it is tested at colliders.

\vspace{0.2cm}
\noindent {\bf Discussion.}
%
CTAO will provide the next leap forward in the search for DM through indirect detection and in this Letter we demonstrate this could result in a near term discovery.
Figure~\ref{fig:TS_over_time} shows that CTAO-North could discover higgsino DM within the current decade and before CTAO-South turns on.
Yet there remain outstanding questions to resolve.
First, more work is needed to understand the shape of the DM density profile in the inner Galaxy.
As seen in Fig.~\ref{fig:J_facs}, the spread in expected $J$-factors in the inner Galaxy currently spans over two orders of magnitude, which translates into a large uncertainty on the expected flux.
Secondly, our results follow from an idealized analysis, with no mismodeling for the backgrounds from misidentified cosmic-rays or astrophysical gamma-rays.
In reality the analysis sensitivity may be further limited by these systematics, particularly that of the cosmic-ray background, which dominates at TeV energies.
As mentioned above, the realistic analysis we envision involves building a robust spectral model for the cosmic-ray background from OFF-data taken at similar observing conditions but away from the Galactic Center.
Preliminary data, which should be available soon from CTAO-North, will help shed light on the viability of this strategy.
Lastly, detecting the higgsino for the true underlying DM profile of the Milky Way may require that both CTAO-North and South successfully install the required instrumentation and allocate sufficient observing resources towards the DM science case.
For the most pessimistic DM profiles, CTAO-South may even need additional instrumentation beyond the Alpha Configuration, which at the moment is not planned.
Higgsino DM, which is one of the best motivated models for one of the most compelling fundamental science questions, provides a strong case for aggressively pursuing such instrumentation.

\vspace{0.2cm}
\noindent {\it Acknowledgments.}
%
We thank Abdelaziz Hussein, Lina Necib, and Andrew Wetzel for discussions related to the DM density in the inner Galaxy. 
This work was conducted in the context of the CTAO Consortium.  We thank the CTAO internal referees for useful comments on the manuscript.
It has made use of the CTAO instrument response functions provided by the CTAO Consortium and Observatory.
This work of SA was supported by JST SPRING, Grant Number JPMJSP2108, and by Grant-in-Aid for JSPS Fellows, Grant Number 24KJ0544.
The work of TI was supported by JSPS KAKENHI Grant Number 20KK0067. 
The work of NLR was supported by the Office of High Energy Physics of the U.S. Department of Energy under contract DE-AC02-05CH11231.
BRS is supported in part by the DOE award DESC0025293, and BRS acknowledges support from the Alfred P. Sloan Foundation. 
The work of WLX was supported by the Kavli Institute for Particle Astrophysics and Cosmology, and by the U.S. Department of Energy under contract DE-AC02-76SF00515.

\bibliographystyle{utphys}
\bibliography{bib}

\clearpage
\appendix
\onecolumngrid
\section*{End Matter}
\twocolumngrid

\setcounter{equation}{0}
\setcounter{figure}{0}
\renewcommand{\theequation}{\thesection\arabic{equation}}
\renewcommand\thefigure{\thesection \arabic{figure}}    

\section{Analysis Details}
\label{app:analysis} 

Here, we provide a more detailed description of the analysis used to construct the projections shown in Fig.~\ref{fig:TS_over_time}.

Consider first the expected signal and background. 
The expected differential flux of gamma-rays from DM annihilation incident on Earth is given by
\begin{equation}
{dN \over dE dt dA} = \frac{J}{8 \pi m_\chi^2} \frac{d \langle \sigma v \rangle}{dE},
\end{equation}
where $d \langle \sigma v \rangle / dE_\gamma$ is the differential cross-section with respect to the photon energy $E$ normalized per annihilation.
For example, if the annihilation proceeds via $\chi \chi \to \gamma\gamma$ with rate $\langle \sigma v \rangle$, we have $d \langle \sigma v \rangle/dE = 2 \langle \sigma v \rangle \delta(E - m_\chi)$.
The $J$-factor is defined as an integral over the DM density~\cite{Bergstrom:1997fh},
\begin{equation}
J(\theta) \equiv \int_0^\infty ds \, \rho_{\scriptscriptstyle {\rm DM}}^2[r(s, \theta) ],
\end{equation}
and the integral is computed along a given direction in the sky with $s$ the distance from Earth.

Given a pointing location on the sky, we forward model the predicted energy- and spatial-dependent flux map to a set of observed counts in (finely) pixelated sky positions and energy bins, using the appropriate CTAO instrument response~\cite{CTAO_2021_5499840}. 
We then re-bin the data into our (coarser) analysis annuli and energy bins, which are used in the projected analysis.
Note that we forward model the {\it Fermi} \texttt{p8r3} diffuse emission in the same way to obtain the expected astrophysical background, per annulus and energy bin.
The expected cosmic-ray-induced background is given directly by the CTAO IRFs~\cite{CTAO_2021_5499840}.

The Asimov data sets are given by the set of predicted counts ${\bm d} = \{N_{i,k} \}$, where $i$ specifies the Galactocentric annulus and $k$ the energy bin.
We analyze these data given the likelihood
\begin{equation}
p({\bm d} | {\mathcal M}, {\bm \theta}) = \prod_{i,k} \frac{\mu_{i,k}({\bm \theta})^{N_{i,k}} e^{-\mu_{i,k}({\bm \theta})}}{N_{i,k} !},
\end{equation}
where for the Asimov (mean, expected) data set we analytically continue the factorial to non-integer counts using the gamma function.
Here, the model ${\mathcal M}$ is the null plus signal model, which has a single model parameter $\mu$, controlling the normalization of the signal flux.
For this idealized projection, we assume perfect knowledge of the background flux in order to separate our results from any assumption of analysis strategy.

We construct the profile likelihood 
\begin{equation}
\lambda(\mu) = \frac{p({\bm d}|{\mathcal M},  \mu)}{p({\bm d}|{\mathcal M}, \hat {\mu})},
\end{equation}
with ${\hat \mu}$ denoting the model parameter point that maximizes the likelihood.
The discovery TS is given by $q \equiv - 2 \ln \lambda(0)$ for $\hat{\mu} > 0$, with $q \equiv 0$ otherwise, since we are interested in a one-sided test.
In the Asimov analysis to determine the expected discovery TS, the data contains a signal component such that $\hat \mu$ always matches the true value of $\mu$ used to create the data.
(See Ref.~\cite{Safdi:2022xkm} for a pedagogical introduction to this framework.)

It's worth mentioning that a realistic analysis scenario will likely contain several nuisance parameters that give freedom to the background model.
In the template analysis we envision, an additional two nuisance parameters per annulus would be included in the likelihood, to be profiled over such that the likelihood is maximized for each choice of $\mu$.
These parameters would control the overall normalizations of the cosmic-ray-induced and astrophysical backgrounds respectively within that spatial region, rescaling all expected counts $\mu_{i,k}$ by the same amount in each energy bin $k$.

We end with a discussion of the relative benefits of conducting an analysis with higgsino annihilation as the specific signal model, as opposed to the more general searches for ``line-only"  $\chi\chi\to \gamma\gamma$ and  ``continuum-only" $\chi\chi\to W^+W^-$ annihilation modes in the gamma-ray data.  
These types of searches are far more common in the literature as they are more model independent in their framing, and each of these signal modes share qualitative features with the higgsino spectrum. 
On the other hand, each is an imperfect partial model of a fully-known and well-motivated signal prediction.
It is worth being concerned that conducting mode-specific and model-independent searches at the exclusion of model-specific ones may result in unnecessarily reduced sensitivity or even mismodeled results.

Under the signal hypothesis, where the Asimov data is generated including flux from a 1.1\,TeV higgsino, we find that the discovery TS is reduced by a factor of $\sim$2 by searching for line-only or $W^+W^-$-only signals.
Since in the Gaussian limit accumulation of TS is linear with time,  twice the amount of observation time would be needed to establish an equal-significance discovery of a line-only signal.
In a realistic analysis the sensitivity of a $W^+W^-$ search would suffer more severely, since the loss of the sharp endpoint feature greatly increases the spectral degeneracy between signal and background.
As a second-order note, the best-fit cross section from a search of $\chi\chi\to \gamma\gamma$ and $\chi\chi \to W^+W^-$ will differ from the ``true" higgsino cross sections.
This is due to end-point contributions to the former, resulting in a best-fit value $\sim$20\% larger, and spectrally different $ZZ$ contributions to the latter, resulting in a best-fit $\sim$40\% smaller than the total annihilation rate.

\newpage
\section{Supplementary Figures}
\label{app:supp_fig}

\begin{figure}[!htb]
\centering
\includegraphics[width=\linewidth]{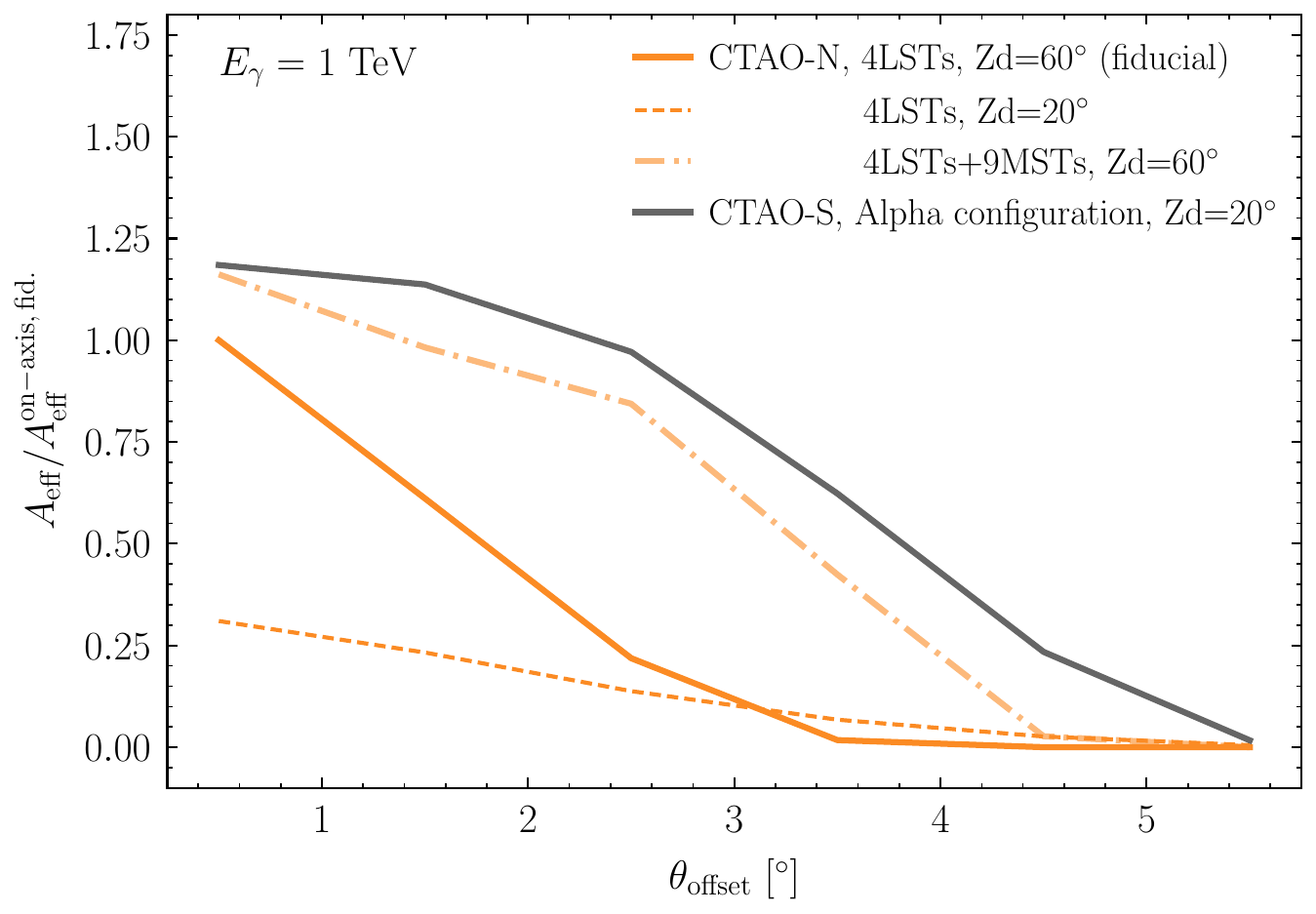}
\includegraphics[width=\linewidth]{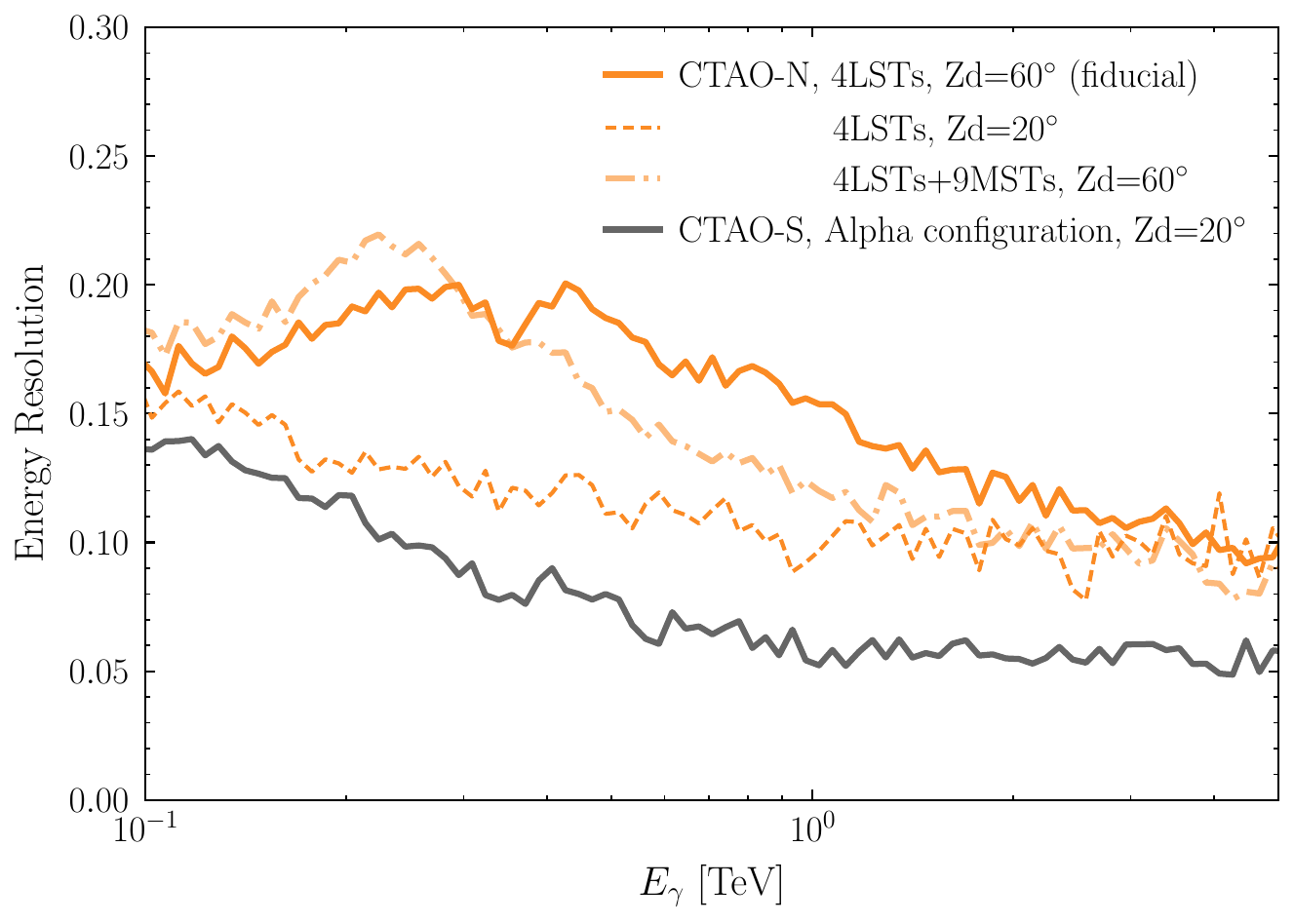}
\includegraphics[width=\linewidth]{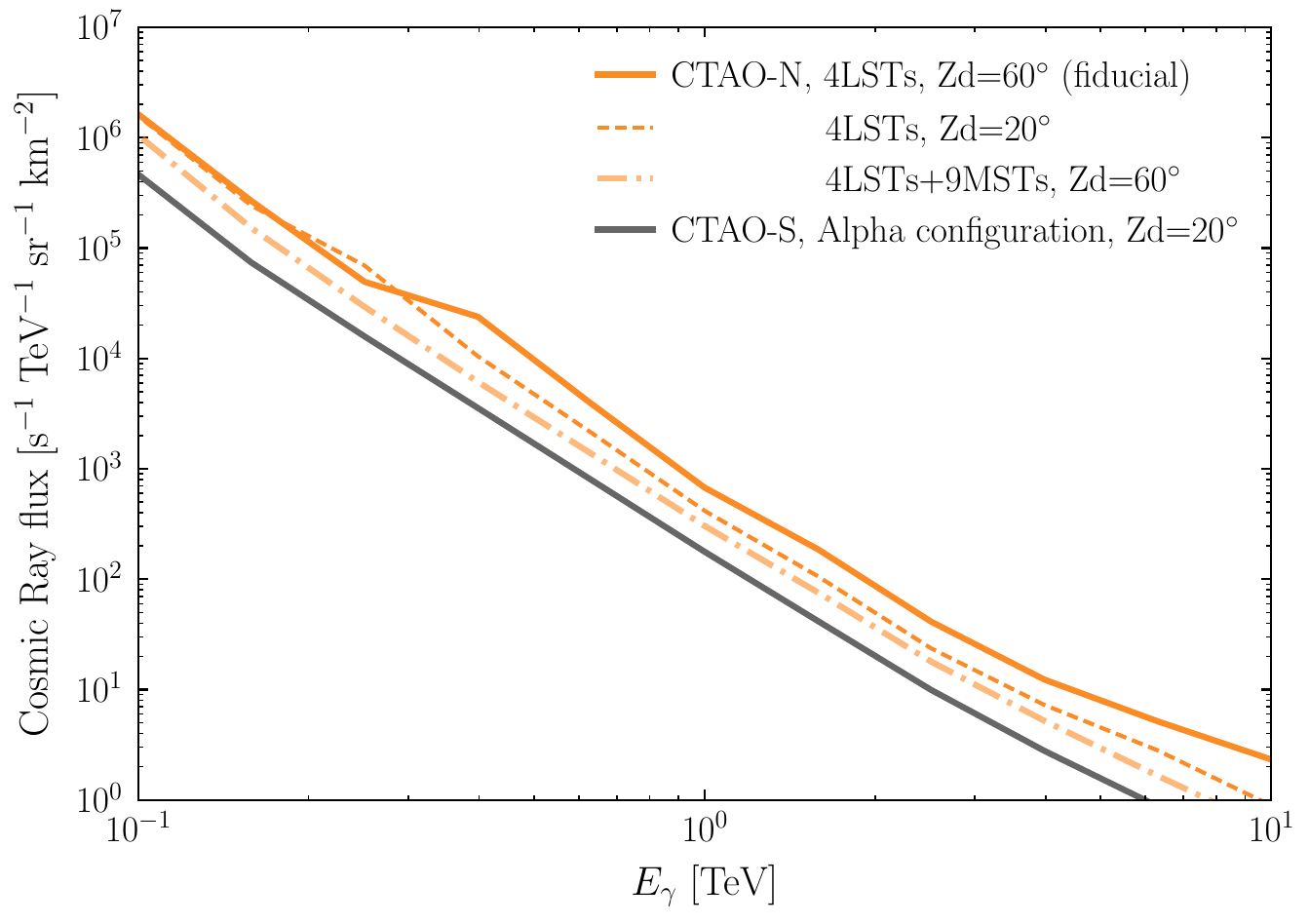}
\vspace{-0.8cm}
\caption{Similar to Fig.~\ref{fig:Aeff}, comparison of various instrumentation metrics amongst the different CTAO instrumental configurations and observations modes considered in the main text.
We show the expected performance of the instrument field-of-view (top), energy resolution (middle) and cosmic-ray background rejection (bottom).
Note that energy resolution is defined here as $\delta E / E$, with $E$ the central line energy and $\delta E$ the 68\% symmetric containment interval on the flux after forward modeling by the energy migration matrix.
}
\vspace{-0.5cm}
\label{fig:comp}
\end{figure}

\begin{figure}[!htb]
\centering
\includegraphics[width=0.8\linewidth]{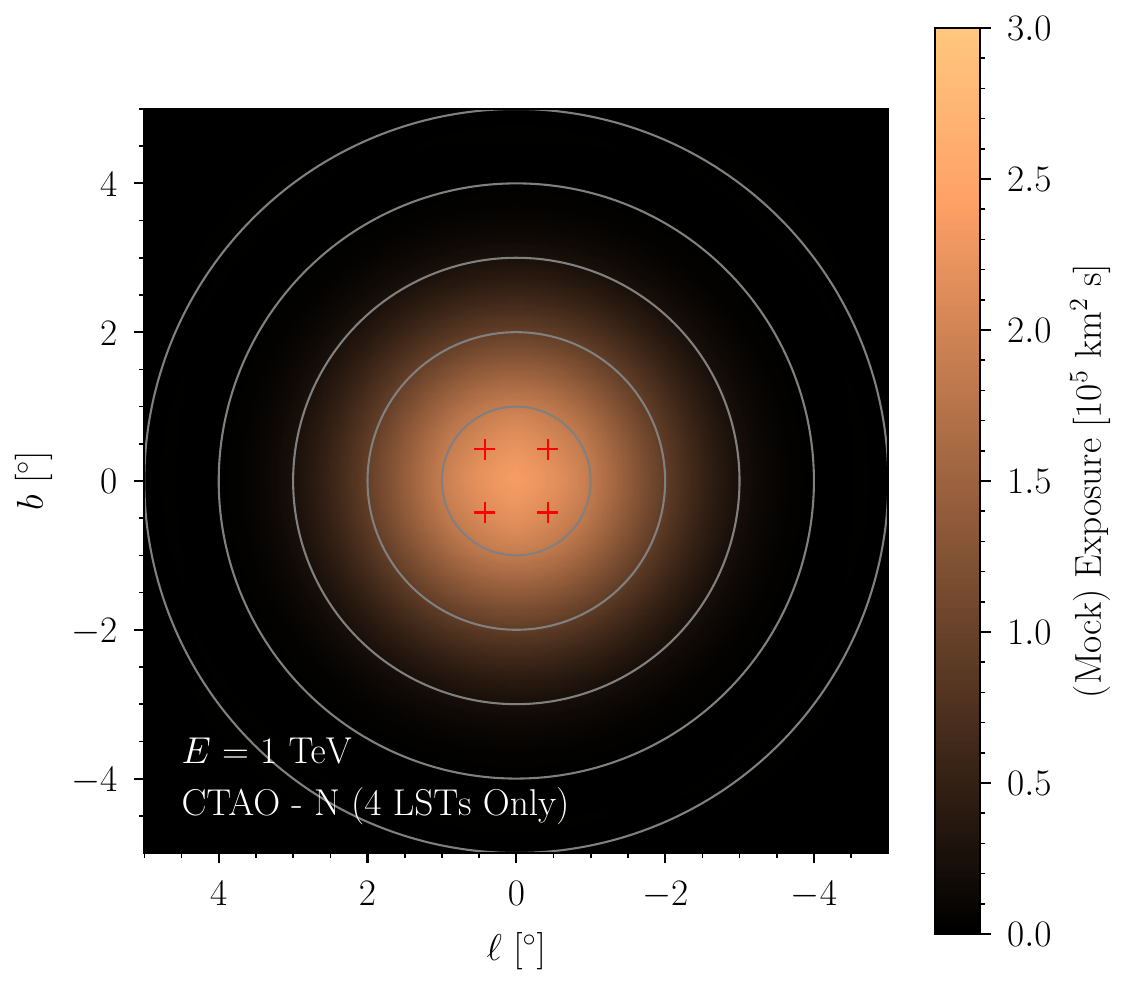}
\includegraphics[width=0.8\linewidth]{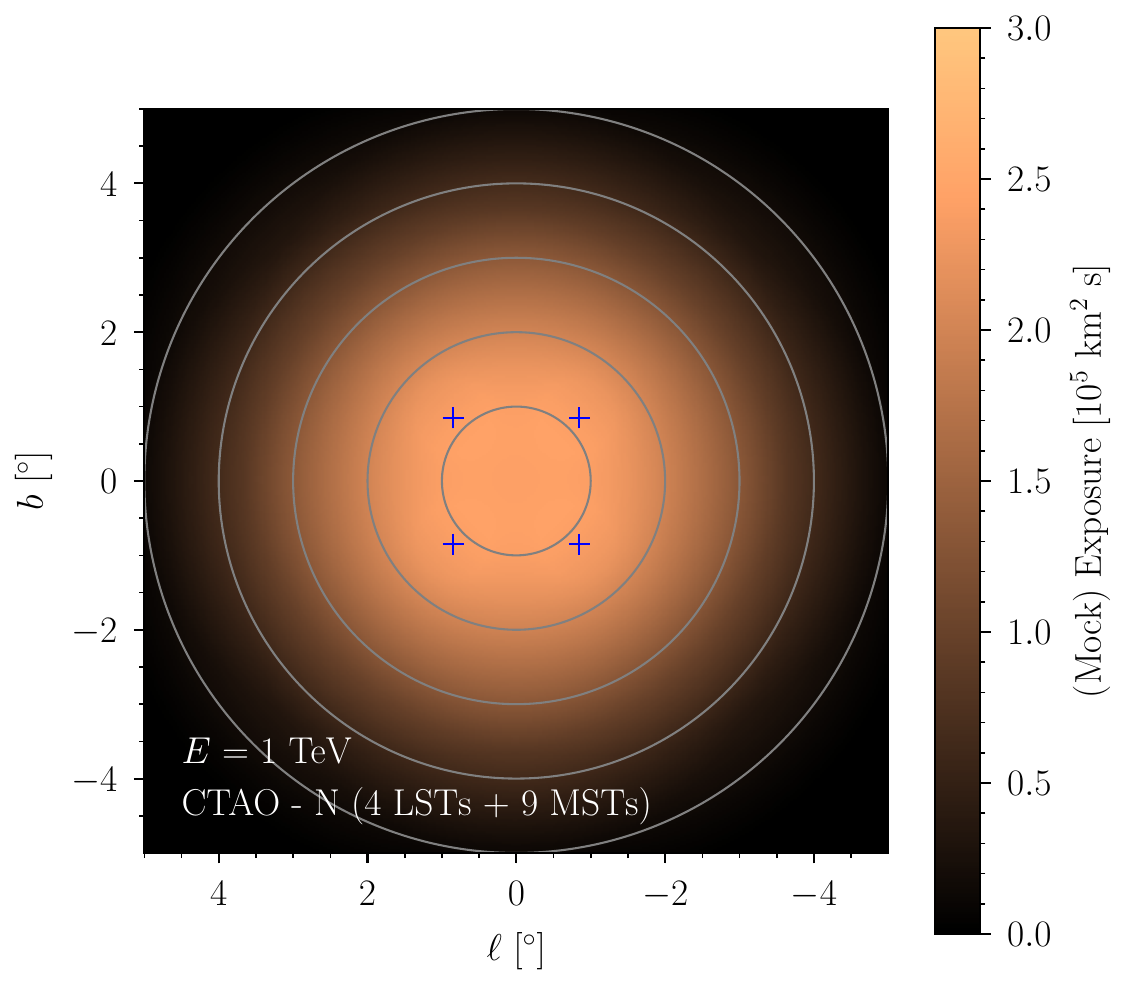}
\includegraphics[width=0.8\linewidth]{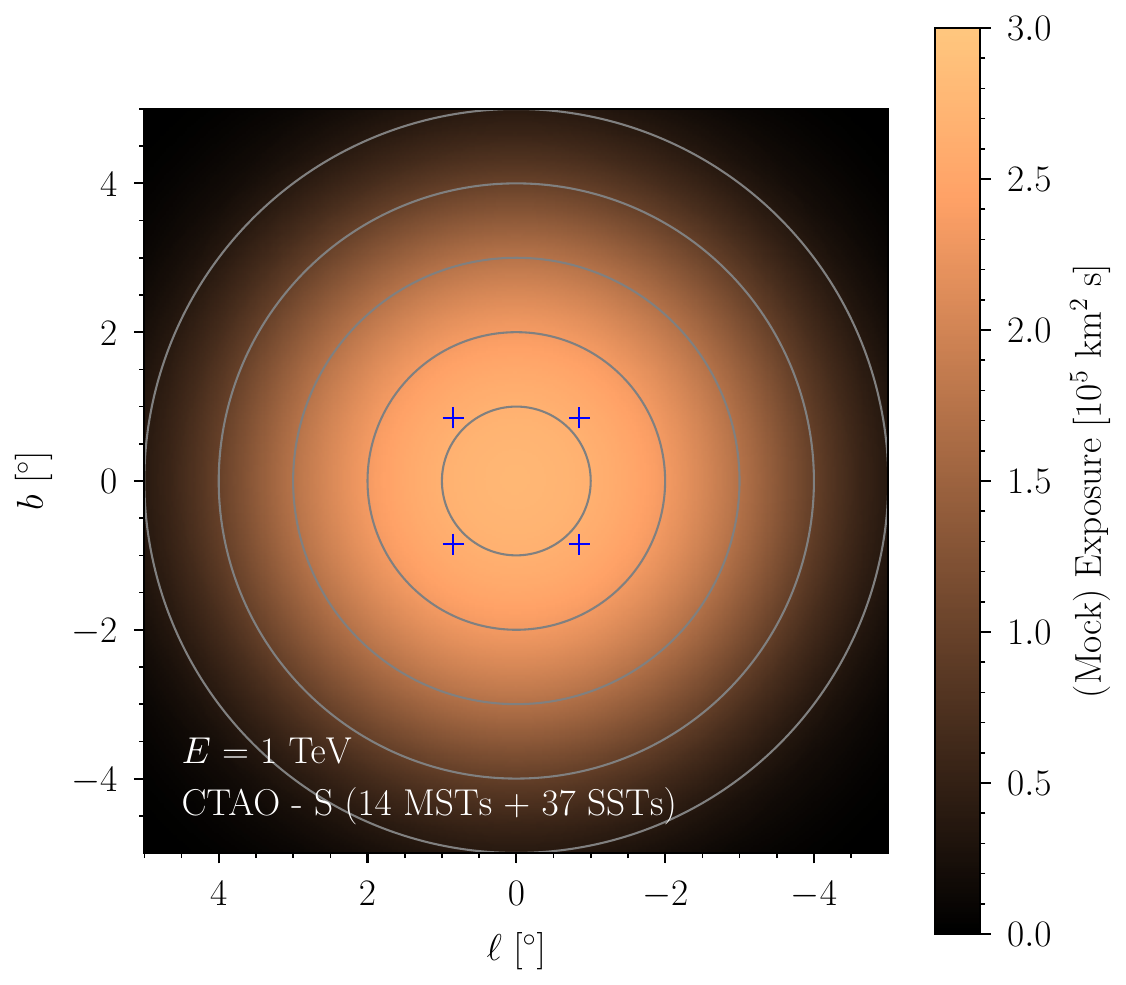}
\vspace{-0.2cm}
\caption{The mock exposure map at 1\,TeV that we generate using Monte Carlo and the assumed instrument response function for our projections of the CTAO-North LST subarray (top), and the full North (middle), and South (bottom) arrays.
We assume pointing locations further off-set from the Galactic Center for the latter configurations due to their larger FOV. 100 hours of total exposure are assumed, equally split between the pointing locations.
}
\vspace{-0.5cm}
\label{fig:exp_map}
\end{figure}

\clearpage

\unappendix

\end{document}